# Efficient high-order analysis of bowtie nanoantennas using the locally corrected Nyström method

Hamid T. Chorsi[1] and Stephen D. Gedney[1,*]

[1]*Department of Electrical Engineering, University of Colorado Denver, Denver, Colorado 80204, USA*
*hamid.chorsi@ucdenver.edu*
[*]*stephen.gedney@ucdenver.edu*

**Abstract:** It is demonstrated that the Locally Corrected Nyström (LCN) method is a versatile and numerically efficient computational method for the modeling of scattering from plasmonic bowtie nanoantennas. The LCN method is a high-order analysis method that can provide exponential convergence. It is straightforward to implement, accurate and computationally efficient. To the best of the author's knowledge, the high-order LCN is here applied for the first time to 3D nanostructures. Numerical results show the accuracy and efficiency of the LCN applied to the electromagnetic analysis of nanostructures.

**1. Introduction**

Optical nanoantennas have generated increasing interest in the past ten years thanks to advances in nanotechnology and nanophotonics. Their unique abilities, such as breaking the diffraction limit of light, confining optical fields to very small dimensions and directly localized optical sources into the far field [1], enable a large number of applications, such as optical imaging [2, 3], optical communications, photovoltaics, optical waveguides, and quantum optics [4, 5]. Among all the nanoparticles, the bowtie structure is attractive because of its triangular geometry, which leads to the "lightning-rod" effects at the gap apexes [6].

Design methods capable of accurately describing the scattering properties of bowtie nanoantennas are essential to the development of optical nanoantenna technology.  The scattering of metallic bowtie nanoantennas has been investigated both theoretically and experimentally using different computational methods in order to study the effect of different bowtie nanoantennas parameters (apex width, bow angle, height and gap separation) [7, 8]. The Finite-difference Time-domain (FDTD) method has been widely used to analyze the scattering properties of metallic nanoparticles [9-12]. In [13] the authors used the FDTD method to investigate Plasmon resonances in silver bowtie nanoantennas.  Authors in [14] investigated the effect of dielectric coating on the optical resonance of plasmonic bowtie nanoantennas using the finite-element method (FEM).  FEM has also been used to analyze Asymmetric Bowtie nano-Colorsorters (ABnC) [15].).  One of the disadvantages of finite methods such as FDTD or FEM is that the entire volume that contains the nanoantenna extending out to an absorbing boundary must be discretized.

The surface integral equation (SIE) method is a popular numerical method used for computing the electromagnetic scattering of 3D objects. Unlike volume integral equation (VIE) methods, SIE methods only require the discretization of the surfaces bounding homogeneous and isotropic domains and thus reduces the number of unknowns compared to methods based on volume discretization.  Previous works to use SIEs in nanophotonics are all based on the well-known method of moments (MoM) using low-order Rao-Wilton-Glisson (RWG) basis and test functions [16-18]. The MoM is complex and typically requires an expensive double integration [19].

Recently, the high-order Locally Corrected Nyström (LCN) method [20] has been introduced as an alternative to the more traditional MoM. The elegance of this technique is in its simplicity and ease of implementation, while it can provide true high-order convergence.  Error can be predicted with p-refinements, eliminating the need for remeshing for convergence studies. A Nyström-type discretization has applied in the literature for 2D nanostrip grating [21].  In this letter, the LCN method [19] is employed for the first time to efficiently analyze the electromagnetic properties of three-dimensional metallic bowtie nanoantennas.  Two different SIEs are studied, namely the Poggio–Miller–Chang–Harrington–Wu–Tsai (PMCHWT) and the Müller methods [22]. The method is first validated for a canonical problem and then applied to the analysis of a 3D bowtie nanoantenna.

## 2. Formulation

Consider the time-harmonic electromagnetic scattering (with $e^{j\omega t}$ time dependence) by a homogeneous dielectric object $V_i$ having a permittivity $\varepsilon_i$ and permeability $\mu_i$ the total electric and magnetic fields in $V_i$ can be expressed as a superposition of incident and scattered fields, where the scattered fields are radiated by the equivalent currents in surface $V_i$ are expressed as [19]:

$$\vec{E}_i^{scat}(\vec{J}_i, \vec{M}_i) = \eta_i \mathbf{L}_i(\vec{J}_i) - \mathbf{K}_i(\vec{M}_i)$$
$$\vec{H}_i^{scat}(\vec{J}_i, \vec{M}_i) = \mathbf{K}_i(\vec{J}_i) + \eta_i^{-1} \mathbf{L}_i(\vec{M}_i) \quad (1)$$

where

$$\mathbf{L}_i(\vec{X}_i) = -jk_i \int_S \left[ \overline{\overline{I}} + \frac{1}{k_i^2} \nabla\nabla \right] \cdot \vec{X}_i(\vec{r}') G_i(\vec{r},\vec{r}') ds',$$
$$\mathbf{K}_i(\vec{X}_i) = \int_S \nabla G_i(\vec{r},\vec{r}') \times \vec{X}_i(\vec{r}') ds', \quad (2)$$

$\vec{J}_i$ and $\vec{M}_i$ are the equivalent currents in $V_i$, $\overline{\overline{I}}$ is the unit dyad, $G_i(\vec{r},\vec{r}') = e^{-jk_i|\vec{r}-\vec{r}'|}/4\pi|\vec{r}-\vec{r}'|$ is the free-space Green's function, $k_i = \omega\sqrt{\varepsilon_i \mu_i}$ is the wave number, and $\eta_i = \sqrt{\mu_i/\varepsilon_i}$ is the characteristic wave impedance in $V_i$. If $S_{i,j}$ is the surface separating two material regions $i$ and $j$, the fields must satisfy the boundary constraints

$$\hat{n}_i \times \vec{E}_i^{inc}\Big|_{S_{i,j}^+} = -\vec{M}_{i,j} - \hat{n}_i \times \vec{E}_i^{scat}\Big|_{S_{i,j}^+}$$
$$-\hat{n}_i \times \vec{E}_j^{inc}\Big|_{S_{i,j}^-} = \vec{M}_{i,j} + \hat{n}_i \times \vec{E}_j^{scat}\Big|_{S_{i,j}^-}$$
$$\hat{n}_i \times \vec{H}_i^{inc}\Big|_{S_{i,j}^+} = \vec{J}_{i,j} - \hat{n}_i \times \vec{H}_i^{scat}\Big|_{S_{i,j}^+} \quad (3)$$
$$-\hat{n}_i \times \vec{H}_j^{inc}\Big|_{S_{i,j}^-} = -\vec{J}_{i,j} + \hat{n}_i \times \vec{H}_j^{scat}\Big|_{S_{i,j}^-}$$

where $\vec{E}^{inc}, \vec{H}^{inc}$ are radiated by impressed sources in $V_i$, $\vec{E}^{scat}, \vec{H}^{scat}$ are radiated by equivalent currents in $V_i$ (similarly in $V_j$), $\hat{n}_i$ is the unit normal to $S_{i,j}$ directed into $V_i$ and $\hat{n}_j = -\hat{n}_i$. Using the boundary conditions in Eq. (3) one can solve for the equivalent electric and magnetic current densities on the material boundaries. Only two constraints are needed to solve for the two unknowns. Using the electric fields or the magnetic fields only on the interior and exterior boundaries will lead to a formulation that can be corrupted by spurious "interior resonances" [22]. A more suitable method would be to scale the interior and exterior constraints separately and then combining them. This can be expressed as [22]

$$\alpha_i \left[ -\hat{n}_i \times \vec{E}_i^{inc}\Big|_{S_{i,j}^+} - \vec{M}_{i,j} - \hat{n}_i \times \vec{E}_i^{scat}\Big|_{S_{i,j}^+} \right] +$$
$$\alpha_j \left[ \hat{n}_i \times \vec{E}_j^{inc}\Big|_{S_{i,j}^-} + \vec{M}_{i,j} + \hat{n}_i \times \vec{E}_j^{scat}\Big|_{S_{i,j}^-} \right] = 0 \quad (4)$$

for the electric field, and

$$\beta_i \left[ -\hat{n}_i \times \vec{H}_i^{inc}\Big|_{S_{i,j}^+} + \vec{J}_{i,j} - \hat{n}_i \times \vec{H}_i^{scat}\Big|_{S_{i,j}^+} \right] + \quad (5)$$
$$\beta_j \left[ \hat{n}_i \times \vec{H}_j^{inc}\Big|_{S_{i,j}^-} - \vec{J}_{i,j} + \hat{n}_i \times \vec{H}_j^{scat}\Big|_{S_{i,j}^-} \right] = 0,$$

for the magnetic field. Among infinitely many possibilities for the scalars $\{\alpha_i, \alpha_j, \beta_i, \beta_j\}$ several choices provide stable formulations, such as

$$\alpha_i = \alpha_j = 1, \qquad \beta_i = \beta_j = 1, \quad (6)$$

and

$$\alpha_i = -\varepsilon_i, \alpha_j = \varepsilon_j, \qquad \beta_i = \mu_i, \beta_j = -\mu_j, \quad (7)$$

which lead to the PMCHWT and Müller formulations, respectively.

In the rest of the paper, numerical examples are presented to demonstrate the validity, accuracy and flexibility of the proposed LCN solution. Details of the LCN discretization of the PMCHWT and Müller formulations can be found in [19, 20, 22, 23]. A mixed-order LCN method [23, 24] was used to appropriately represent the current density. High-order surface discretization [19] is also used.

## 3. Simulation results and discussion

For validation purposes, the scattering by a sphere is considered since the LCN solution can be compared to the analytical solution based on a Mie series. A gold nanosphere with a radius of 200 nm is illuminated with a vertically polarized plane wave at 467.7644 THz ($\lambda = 641$ nm) incident from $(\theta^{inc}, \phi^{inc}) = (0°, 0°)$. At this wavelength, the relative dielectric constant of gold is $\varepsilon_r = -11.095 - j1.2603$ [25]. The sphere was discretized with 64, fourth-order quadrilateral cells. Figure 1. shows the bi-static radar cross section (RCS) patterns of the VV ($\theta - \theta$) and HH ($\phi - \phi$) polarizations computed via the LCN method with a second-order discretization of the current densities (p=2) for the PMCHWT and Müller formulations. The results are compared with an analytical Mie-series solution. All three results are indistinguishable on the graph.

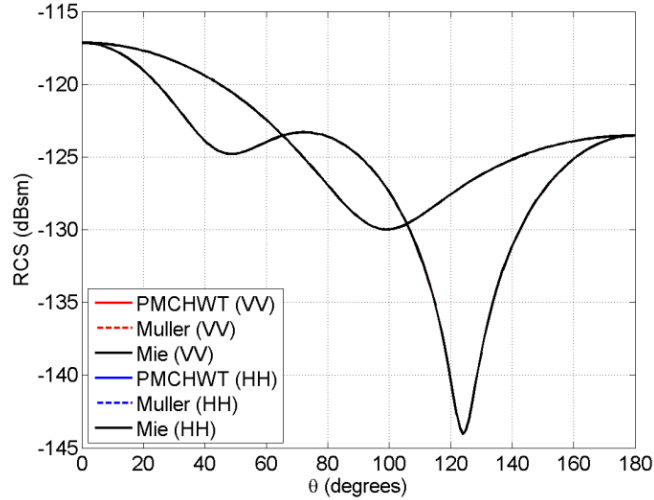

Fig. 1. Comparison of the LCN solution (p=2) via the PMCHWT and Müller formulations with the analytical Mie series solution for the 200 nm radius gold nanosphere.

Table 1: Comparison of resources and errors for the LCN simulation of the 200 nm radius nanosphere at 467.7644 THz.

| Method | # Cells | p | N | Error(VV) | Error(HH) | CPU-sec (Fill/Factor) |
|---|---|---|---|---|---|---|
| PMCHWT | 64 | 0 | 512 | 1.547e-1 | 1.646e-1 | 2.9 s/0.014 s |
| PMCHWT | 64 | 1 | 1536 | 4.531e-3 | 4.614e-3 | 5.27 s / 0.24 s |
| PMCHWT | 64 | 2 | 3072 | 1.974e-3 | 1.996e-3 | 8.629 s / 0.7 s |
| Müller | 64 | 0 | 512 | 1.913e-1 | 1.734e-2 | 1.59 s/ 0.087 s |
| Müller | 64 | 1 | 1536 | 3.601e-3 | 2.173e-3 | 3.39 s / 1.94 s |
| Müller | 64 | 2 | 3072 | 1.488e-4 | 1.406e-4 | 6.56 s / 9.24 s |

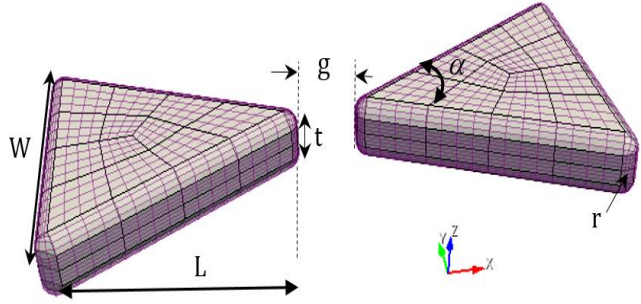

Fig. 2. Geometry of the bowtie nanoantennas meshed with a total of 226 quadratic quadrilateral cells. g is the gap size, α is the bow or gap angle, t is the hight, L is the length, W is the antenna width, and r is the rounded corner radii.

Table 1 illustrates the error in the bi-static RCS relative the Mie-series solution as a function of the LCN order. In this table, *p* is the order of the mixed-order LCN basis. N is the total number of unknowns (representing both J and M). The errors in the RCS, relative to the Mie series solution and averaged over the unit sphere, are given. CPU times for computing (fill) and solving (factor) the system matrix are also provided. These were recorded on an Intel Xeon E5-1650 (3.5 GHz) CPU with 12 threads. From Table 1 it is observed that increasing the order of the mixed-order basis exhibits exponential convergence. However, for the sphere, the Müller method is converging faster. This is because the Müller method has a reduced singularity and the LCN method represents the diagonal operator exactly [23].

Next, the LCN method is applied to the scattering of a bowtie nanoantenna. The structure and the geometric parameters of the 3D bowtie nanoantenna are presented in Fig 2. The bowtie has rounded corners with a radius of curvature r. The bowtie is also assumed to be composed of a homogeneous, isotropic material which is gold. For sake of illustration, a bowtie structure with the dimensions L = 50 nm, W = 50 nm, t = 10 nm, g =11.98 nm, r = 5 nm, α= 53.13˚, is considered.

Figure 3. Shows the bi-static RCS of the presented bowtie nanoantenna when it is illuminated with vertically and horizontally polarized plane waves with $(\theta^{inc}, \phi^{inc}) = (0°, 0°)$ at a frequency of f = 467.7644 THz $(\lambda = 641 \text{ nm})$, assuming the relative permittivity of the gold to be $\varepsilon_r = -11.095 - j1.2603$ at this wavelength [25].

The surface of each half of the bowtie-antenna was discretized with 226 quadratic quadrilateral cells, as illustrated in Fig. 2. The LCN solution was obtained using both the Müller and the PMCHWT formulations for 0th-order, 1st-order, 2nd-order and 3rd-order mixed-order LCN discretizations. The resulting bi-static RCS is illustrated in Fig. 3 for V-V and H-H polarizations in the $\phi = 0°$ plane. (Note that the optical cross section (OCS) is the backscatter angle $\theta = 0°$.) Both the Müller and PMCHWT formulations have converged to 3 digits of accuracy for the 1st-order (p=1) discretization. The magnitude of the surface current densities on the bowtie surface for the vertically polarized incident wave is illustrated in Fig. 4. In this figure, it is observed that the light is strongly focused in the gap region.

The resonance of the bowtie antenna in Fig. 2 is studied. To this end, the bowtie antenna was illuminated by a normally incident vertically polarized plane wave. The frequency of the

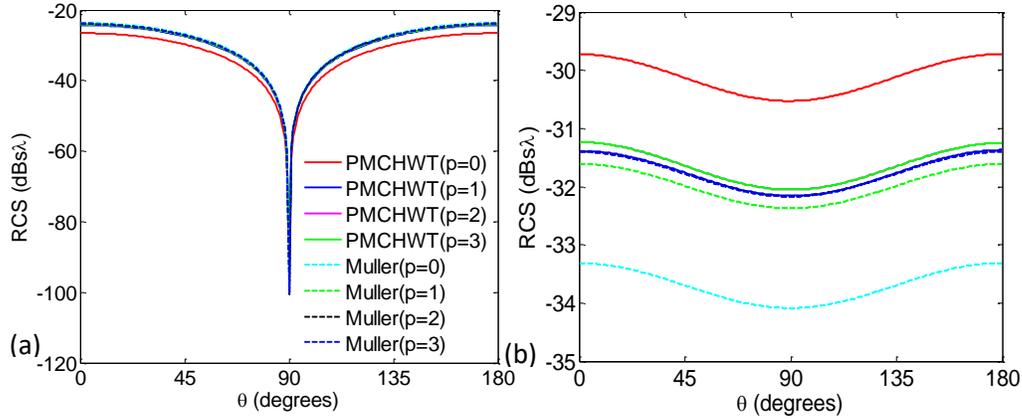

Fig. 3. Bi-static RCS of the bowtie in the $\phi = 0°$ plane with f = 467.7644 THz ($\lambda = 641$ nm), $(\theta^{inc}, \phi^{inc}) = (0°, 0°)$, PMCWT and Müller formulation, for a discretization with 226 quadratic quadrilateral cells (total), LCN solution. (a) $\sigma_{V-V}$, (b) $\sigma_{H-H}$.

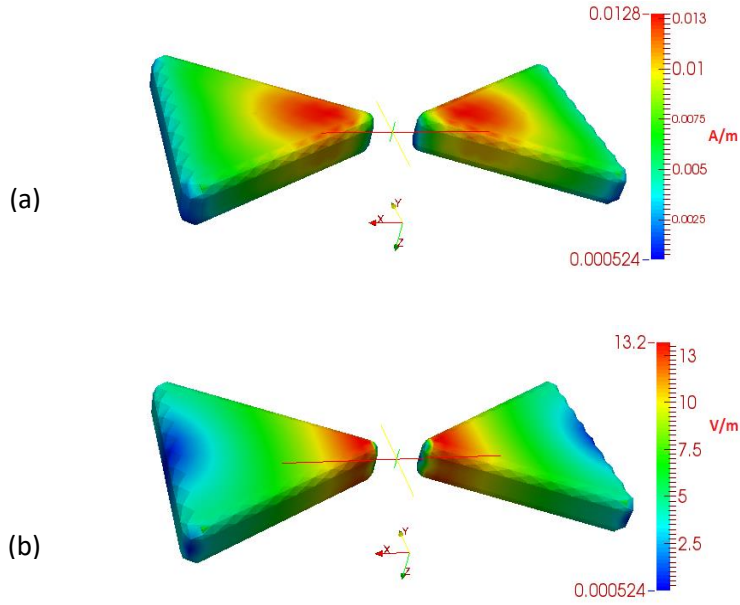

Fig. 4. Equivalent electric current (a) and equivalent magnetic current (b) induced on the bowtie computed using Müller's method, at f = 467.7644 THz ($\lambda = 641$ nm) due to a normally incident vertically polarized plane wave ($\vec{E}^{inc} = \hat{x} e^{jk_0 z}$ V/m).

plane wave was swept from 350 THz to 750 THz. The field enhancement in the gap of the bowtie was computed as

$$f_E = \int_{gap} \vec{E}^{scat} \cdot d\vec{\ell} \qquad (8)$$

through the center of the gap region. The frequency dependence of the relative permittivity of the gold was computed according to [25]. This was simulated using the Müller method for the mesh in Fig. 2 and with first-order mixed-order basis (p=1). The results are illustrated in Fig. 5. The field enhancement peaks at 467.7644 THz.

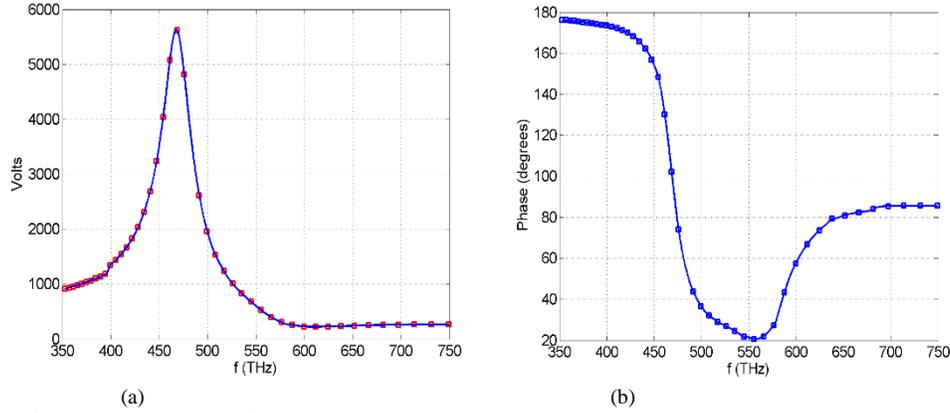
(a)                                 (b)
Fig. 5. Plasmonic resonance of the bowtie nanoantenna, magnitude (a), and phase (b), of the electric field enhancement (8).

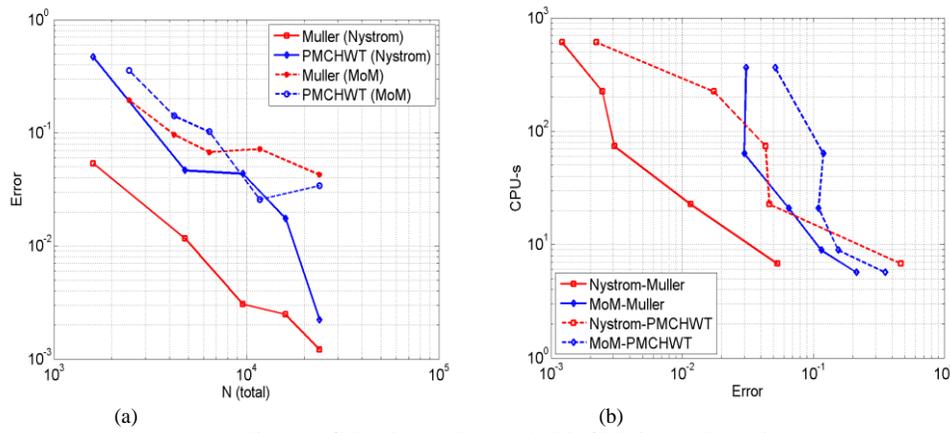
(a)                                 (b)
Fig. 6 . LCN vs. MoM, Error vs. N (a), CPU time vs. Error (b).

Figure 6 shows a plot of the error resulting from both the LCN and a MoM solution with respect to the number of unknowns for the bowtie nanoantenna at the resonance frequency (467.7644 THz). The MoM solution used standard RWG basis on flat faceted triangles. Also shown in Fig. 6 is the CPU-time versus error for both methods. The LCN mesh consisted of 226, fourth-order quadrilateral cells. The order was increased from 0th to 4th order. The MoM mesh density was increased from 656 triangles at the coarsest to 5980 triangles at the finest. A dramatic improvement in the convergence and efficiency of the LCN solution is observed as compared to the MoM solution. The MoM solution error is stagnated since the density of faceted traingles is still not sufficient to resolve the rounded corners. The advantage of the LCN is particularly pronounced when a higher accuracy is desired.

### 7. Summary
In summary, it has been shown that the LCN method is an accurate and efficient method for simulating nanoantenna structures. Both PMCHWT and Müller methods were studied. It was found that for smooth structures, the Müller converges at a faster rate. The reason for this is that the Müller has a reduced singularity in the kernel [23]. For structures with sharp edges, the PMCHWT method can be more accurate [23]. It is also shown that the exponential convergence of the LCN method provides an efficient means to control the accuracy of the simulation, and is a viable method for simulating complex nanoantenna structures. Through the use of a layered Green's function, the presented method can also be applied to nano-structures in a layered media, which will be considered in a future work.

## References

1. D. Solís, J. Taboada, M. Araújo, F. Obelleiro, and J. Rubiños-López, "Design of optical wide-band log-periodic nanoantennas using surface integral equation techniques," Opt. Comm. **301**, 61-66 (2013).
2. Höppener, and L. Novotny, "Antenna-based optical imaging of single Ca2+ transmembrane proteins in liquids," Nano. Lett. **8**, 642-646 (2008).
3. Bouhelier, M. Beversluis, and L. Novotny, "Applications of field-enhanced near-field optical microscopy," Ultramicroscopy **100**, 413-419 (2004).
4. L. Novotny, and N. Van Hulst, "Antennas for light," Nat. Photo. **5**, 83-90 (2011).
5. S. A. Maier, P. G. Kik, and H. A. Atwater, "Optical pulse propagation in metal nanoparticle chain waveguides," Phys. Rev. B **67**, 205402 (2003).
6. R. D. Grober, R. J. Schoelkopf, and D. E. Prober, "Optical antenna: Towards a unity efficiency near-field optical probe," J. Appl. Phys. **70**, 1354-1356 (1997).
7. E. X. Jin, and X. Xu, "Enhanced optical near field from a bowtie aperture," Appl. Phys. Lett. **88**, 153110 (2006).
8. D. P. Fromm, A. Sundaramurthy, P. J. Schuck, G. Kino, and W. Moerner, "Gap-dependent optical coupling of single "bowtie" nanoantennas resonant in the visible," Nano. Lett. **4**, 957-961 (2004).
9. Sundaramurthy, K. Crozier, G. Kino, D. Fromm, P. Schuck, and W. Moerner, "Field enhancement and gap-dependent resonance in a system of two opposing tip-to-tip Au nanotriangles," Phys. Rev. B **72**, 165409 (2005).
10. N. Felidj, J. Grand, G. Laurent, J. Aubard, G. Levi, A. Hohenau, N. Galler, F. Aussenegg, and J. Krenn, "Multipolar surface plasmon peaks on gold nanotriangles," J. Chem. Phys., **128**, 094702 (2008).
11. E. Hao, and G. C. Schatz, "Electromagnetic fields around silver nanoparticles and dimers," J. Chem. Phys., **120**, 357-366 (2004).
12. Şendur, and W. Challener, "Near-field radiation of bow-tie antennas and apertures at optical frequencies," J. Microsc. **210**, 279-283 (2003).
13. W. Ding, R. Bachelot, S. Kostcheev, P. Royer, and R. E. de Lamaestre, "Surface plasmon resonances in silver Bowtie nanoantennas with varied bow angles," J. Appl. Phys. **108**, 124314 (2010).
14. T.-R. Lin, S.-W. Chang, S. L. Chuang, Z. Zhang, and P. J. Schuck, "Coating effect on optical resonance of plasmonic nanobowtie antenna," Appl. Phys. Lett. **97**, 063106 (2010).
15. Z. Zhang, A. Weber-Bargioni, S. Wu, S. Dhuey, S. Cabrini, and P. J. Schuck, "Manipulating nanoscale light fields with the asymmetric bowtie nano-colorsorter," Nano. Lett. **9**, 4505-4509 (2009).
16. J. M. Taboada, J. Rivero, F. Obelleiro, M. G. Araújo, and L. Landesa, "Method-of-moments formulation for the analysis of plasmonic nano-optical antennas," J. Opt. Soc. Am. A **28**, 1341-1348 (2011).
17. M. Araújo, J. Taboada, D. Solís, J. Rivero, L. Landesa, and F. Obelleiro, "Comparison of surface integral equation formulations for electromagnetic analysis of plasmonic nanoscatterers," Opt. Express **20**, 9161-9171 (2012).
18. Forestiere, G. Iadarola, G. Rubinacci, A. Tamburrino, L. Dal Negro, and G. Miano, "Surface integral formulations for the design of plasmonic nanostructures," J. Opt. Soc. Am. A **29**, 2314-2327 (2012).
19. D. Gedney, and J. C. Young, "*The Locally Corrected Nyström Method for Electromagnetics*" (Springer, 2014), Chap. 5.
20. L. F. Canino, J. J. Ottusch, M. A. Stalzer, J. L. Visher, and S. M. Wandzura, "Numerical solution of the Helmholtz equation in 2D and 3D using a high-order Nyström discretization," J. Comput. Phys. **146**, 627-663 (1998).
21. O. V. Shapoval, J. Ctyroky, and A. I. Nosich, "Mathematical simulation of optical nanoantenna based on a comb-like finite nanostrip grating," *in 2013 IEEE XXXIII International Scientific Conference Electronics and Nanotechnology* (ELNANO)(2013).
22. R. F. Harrington, "Boundary integral formulations for homogeneous material bodies," J. electromagnetic waves and applications **3**, 1-15 (1989).
23. A. Zhu, S. D. Gedney, and J. L. Visher, "A study of combined field formulations for material scattering for a locally corrected Nyström discretization," IEEE Trans. Antennas Propag. **53**, 4111-4120 (2005).
24. S. D. Gedney, A. Zhu, and C.-C. Lu, "Study of mixed-order basis functions for the locally corrected Nyström method," IEEE Trans. Antennas Propag. **52**, 2996-3004 (2004).
25. P. B. Johnson, and R.-W. Christy, "Optical constants of the noble metals," Phys. Rev. B **6**, 4370 (1972).